\documentclass[aps,superscriptaddress,showpacs,nofootinbib]{revtex4}
\usepackage{amssymb}
\usepackage{graphicx}
\begin{document}
%-------------------------------------------------------------------------
% Definition needed for the heading
%-------------------------------------------------------------------------
\def\Barcelo{Barcel\'o}
\def\Schrodinger{Schr\"odinger}
%-------------------------------------------------------------------------
\title[Semiclassical analogue gravity in BEC\dots]
{Probing semiclassical analogue gravity in Bose--Einstein condensates\\ 
with widely tunable interactions}
%-------------------------------------------------------------------------
\author{Carlos \Barcelo}
\email[]{carlos.barcelo@port.ac.uk}
\homepage[http://www.tech.port.ac.uk/staffweb/barceloc/]{}
%\thanks{}
\affiliation{Institute of Cosmology and Gravitation,
University of Portsmouth, Portsmouth PO1 2EG, Britain}
%-------------------------------------------------------------------------
\author{S. Liberati}
\email[]{liberati@physics.umd.edu}
%\homepage[http://www.sissa.it/\~{}liberati]{}
\homepage[http://www2.physics.umd.edu/\~{}liberati]{}
%\thanks{}
\affiliation{Physics Department, University of Maryland, College Park,
MD 20742--4111, USA} 
%-------------------------------------------------------------------------
\author{Matt Visser}
\email[]{Matt.Visser@mcs.vuw.ac.nz}
\homepage[http://www.mcs.vuw.ac.nz/\~{}visser]{}
%\thanks{}
%\altaffiliation{}
\affiliation{School of Mathematical and Computing Sciences,
Victoria University of Wellington, New Zealand}
%-------------------------------------------------------------------------

%-------------------------------------------------------------------------
% Author-specific definitions
%-------------------------------------------------------------------------
\def\g{\kappa}
\def\half{{1\over2}}
\def\L{{\mathcal L}}
\def\S{{\mathcal S}}
\def\d{{\mathrm{d}}}
\def\x{{\mathbf x}}
\def\v{{\mathbf v}}
\def\k{{\mathbf k}}
\def\im{{\rm i}}
\def\etal{{\emph{et al\/}}}
\def\det{{\mathrm{det}}}
\def\tr{{\mathrm{tr}}}
\def\ie{{\emph{i.e.}}}
\def\bnabla{\mbox{\boldmath$\nabla$}}
\def\Box{\kern0.5pt{\lower0.1pt\vbox{\hrule height.5pt width 6.8pt
    \hbox{\vrule width.5pt height6pt \kern6pt \vrule width.3pt}
    \hrule height.3pt width 6.8pt} }\kern1.5pt}
\def\HRULE{{\bigskip\hrule\bigskip}}
%-------------------------------------------------------------------------
\def\be{\begin{equation}}
\def\ee{\end{equation}}
\def\implies{\Rightarrow}
\def\sech{{\mathrm{sech}}}
%-------------------------------------------------------------------------

%-------------------------------------------------------------------------
\date{21 July 2003; Update 14 Aug 2003; \LaTeX-ed \today}
%-------------------------------------------------------------------------
\begin{abstract}
%-------------------------------------------------------------------------
  
  Bose-Einstein condensates (BEC) have recently been the subject of
  considerable study as possible analogue models of general
  relativity.  In particular it was shown that the propagation of
  phase perturbations in a BEC can, under certain conditions, closely
  mimic the dynamics of scalar quantum fields in curved spacetimes. In
  two previous articles [gr-qc/0110036, gr-qc/0305061] we noted that a
  varying scattering length in the BEC corresponds to a varying speed
  of light in the ``effective metric''.  Recent experiments have
  indeed achieved a controlled tuning of the scattering length in
  Rubidium 85.  In this article we shall discuss the prospects for the
  use of this particular experimental effect to test some of the
  predictions of semiclassical quantum gravity, for instance, particle
  production in an expanding universe.  We stress that these effects
  are generally much larger than the Hawking radiation expected from
  causal horizons, and so there are much better chances for their
  detection in the near future.

%\vskip 30pt
%\noindent
%Version 1.0; file {\sf wti-final.tex}

%-------------------------------------------------------------------------
\end{abstract}
%-------------------------------------------------------------------------
\pacs{04.70.Dy, 03.75.Fi, 04.80.-y, cond-mat/0307491}
%-------------------------------------------------------------------------
\widetext
\maketitle
%-------------------------------------------------------------------------
%-------------------------------------------------------------------------
% Author-specific definitions
%-------------------------------------------------------------------------
\def\g{\kappa}
\def\half{{1\over2}}
\def\L{{\mathcal L}}
\def\S{{\mathcal S}}
\def\d{{\mathrm{d}}}
\def\x{{\mathbf x}}
\def\v{{\mathbf v}}
\def\im{{\rm i}}
\def\etal{{\emph{et al\/}}}
\def\det{{\mathrm{det}}}
\def\tr{{\mathrm{tr}}}
\def\ie{{\emph{i.e.}}}
\def\bnabla{\mbox{\boldmath$\nabla$}}
\def\Box{\kern0.5pt{\lower0.1pt\vbox{\hrule height.5pt width 6.8pt
    \hbox{\vrule width.5pt height6pt \kern6pt \vrule width.3pt}
    \hrule height.3pt width 6.8pt} }\kern1.5pt}
\def\HRULE{{\bigskip\hrule\bigskip}}
%-------------------------------------------------------------------------

%----------------------------------------------------------------

%----------------------------------------------------------------
\section{Introduction --- Motivation}
%----------------------------------------------------------------

Semiclassical gravity has played a central role in theoretical
physics.  Phenomena such as the Hawking effect or cosmological
particle production are commonly considered to be crucial first steps
on the way to building up a consistent fully-quantum theory of gravity
(see for example \cite{Birrell-Davies}).  However a fundamental limit
to these investigations is imposed by the fact that their most basic
description is based on linear QFT on a fixed (classical) continuum
spacetime. Several theoretical approaches have been developed to
overcome this limitation: In a fashion that we can call ``top-down'',
string models [brane models] have in some special situations developed
a high energy description of the Hawking
effect~\cite{Vafa-Strominger}, while ``bottom-up'' approaches, based
on Stochastic gravity and the Einstein--Langevin analysis of particle
creation by a gravitational field, have in recent years provided
further insight \cite{Hu1,Hu2}.

On the other hand, the physics community has so far lacked any
possibility for direct experimental tests of these ideas. Indeed this
lack of experimental guidance is a severe hindrance with respect to
further developments in semiclassical gravity (or full-fledged quantum
gravity for that matter). In particular we have no \emph{experimental}
guidance regarding the manner in which the predictions of curved
spacetime quantum field theory are changed once the hypotheses of
non-discreteness and/or a non-fluctuating background are relaxed.  In
this regard the analogue models of gravity developed in recent years
can be considered a first attempt to create an arena which can serve
as a theoretical, and possibly observational, laboratory to test
aspects of these scenarios.

No experimental set up has yet been realized in which the predictions
of analogue models can be observationally tested. Nevertheless
theoretical analyses of analogue models~\cite{Unruh,acoustic} have
been so far remarkably successful in teaching us how semiclassical
gravity phenomena are sensitive to possible quantum gravity effects,
such as, for example, modified (Lorentz violating) dispersion
relations~\cite{Parentani}.  (See for example the trans-Planckian
problem in the Hawking effect \cite{Unruh} and in cosmology
\cite{Brandenberger}.)

What we intend to discuss in this article is a particular class of
experiment ---that we hope could be realized in the very near
future--- wherein certain analogue gravity model predictions could be
tested.  The interest in doing so would not just be that of confirming
a now well-established theoretical prediction, but mainly trying to
evince deviations from the naive theoretical predictions due to the
intrinsic discrete nature of the experimental system and/or to the
possible role of non-linearities.

We shall focus our attention on the analogue gravity system
established by the propagation of linearized phase perturbations in a
Bose-Einstein condensate~\cite{Garay1,Garay2,bec-cqg,abh,laval,grf}.
In particular we shall consider an experiment where a time-varying
scattering length is used to simulate the cosmological expansion of
the universe, and its associated quantum creation of particles.

It is interesting to note that in reference~\cite{Hu-Calzetta} the
authors proposed an explanation of the so called ``bosenova''
phenomenon~\cite{bosenova-exp} (a controlled instability of the bulk
condensate induced by a sign variation of the scattering length)
through a particular implementation of a version of analog
cosmological particle production. In that approach the entire bulk of
the condensate is rendered unstable and suffers catastrophic
breakdown.  Our current paper takes a complementary approach: Instead
of trying to explain an observed phenomenon via analog cosmological
particle production, we shall instead consider the most favourable
conditions to observe it --- preferably without violent disruption of
the entire condensate.

The scheme of the paper will be as follows: In the next Section we
will review the physics of BECs regarding its analogue gravity
features.  Section III will be devoted to the discussion of how to
simulate a FRW universe within a BEC. There exist two main routes to
do this. One considers an explosive expansion of the condensate; the
other makes use of the possibility of tuning the strength of the
interaction between the different bosons in the condensate. This
latter route will be the main concern of this paper.

In Section IV we will first qualitatively describe how the
modification of the interaction strength (encoded in the value of the
scattering length) yields cosmological particle creation. Next, in
Subsection A we will discuss [in the context of current BEC
technology] whether there exists a regime in which this particle
creation process can actually be reproduced. We will see that there is
a limit on the rapidity of change of the background configuration,
associated with the need to enforce a ``Markov approximation'', in
order for the whole construction to make sense. However, this bound
still leaves a lot of parameter space available to look for the
particle creation effect. Subsection B reviews the cosmological
particle creation process, emphasizing the particular features of BEC
systems. Then, Subsection C discuss the actual observability of the
effect. Finally, we conclude with a short summary and discussion.

%----------------------------------------------------------------------
\section{Analogue gravity in Bose--Einstein condensates}
%---------------------------------------------------------------------

Bose-Einstein condensates (BEC) have recently become subject of
extensive study as possible analogue models of general relativity
\cite{Garay1,Garay2,bec-cqg,abh,laval,grf}. In particular it was shown
that the propagation of phase perturbations in a BEC can under certain
conditions closely mimic the dynamics of quantum fields in curved
spacetimes. In previous papers we noted that a varying scattering
length in the BEC system corresponds to a varying speed of light in
the ``effective metric'' \cite{laval,grf}.  Recent experiments have
indeed achieved a controlled tuning of the scattering length in
Rubidium 85 \cite{Rb1}.  The effect is powerful enough to lead to
large non-perturbative changes in the effective metric. Let us start
by very briefly reviewing the derivation of the acoustic metric for a
BEC system.

In the dilute gas approximation, one can describe a Bose gas through a
quantum field ${\widehat \Psi}$ satisfying
\begin{eqnarray}
 \im \hbar \; \frac{\partial }{\partial t} {\widehat \Psi}=
 \left( - {\hbar^2 \over 2m} \nabla^2  + 
 V_{\rm ext}(\x) 
 +\g(a){\widehat \Psi}^{\dagger}{\widehat \Psi}  \right){\widehat \Psi}.
\end{eqnarray}
Here $\g$ parameterizes the strength of the interactions between the
different bosons in the gas. It can be re-expressed in terms of the
scattering length as
\begin{eqnarray}
 \g(a) = {4\pi a \hbar^2\over m}.
\end{eqnarray}
As usual, the quantum field can be separated into a macroscopic
(classical) condensate and a fluctuation: ${\widehat
  \Psi}=\psi+{\widehat \varphi}$, with $\langle {\widehat \Psi}
\rangle=\psi $. Then, by adopting the self-consistent mean field
approximation \cite{Griffin}
\begin{eqnarray}
{\widehat \varphi}^{\dagger}{\widehat \varphi}{\widehat \varphi} \simeq
2\langle {\widehat \varphi}^{\dagger}{\widehat \varphi} \rangle
{\widehat \varphi}
+ \langle {\widehat \varphi} {\widehat \varphi} \rangle 
{\widehat \varphi}^{\dagger},
\end{eqnarray}
one can arrive at the set of coupled equations:
\begin{eqnarray}
 \im \hbar \; \frac{\partial }{\partial t} \psi(t,\x)= \left (
 - {\hbar^2 \over 2m} \nabla^2 
 + V_{\rm ext}(\x)
 + \g \; n_c \right) \psi(t,\x)
+ \g \left(2\tilde n  \psi(t,\x)+ \tilde m \psi^*(t,\x) \right);
\end{eqnarray}
\begin{eqnarray}
 \im \hbar \; \frac{\partial }{\partial t} {\widehat \varphi}=
 \left( - {\hbar^2 \over 2m} \nabla^2  + 
 V_{\rm ext}(\x) 
 +\g \;2 n_T \right){\widehat \varphi} +
 \g \; m_T \; {\widehat \varphi}^{\dagger}.
 \label{quantum-field}
\end{eqnarray}
Here 
\begin{eqnarray}
&&n_c \equiv \left| \psi(t,\x) \right|^2; 
\hspace{2cm}
m_c \equiv \psi^2(t,\x); \\
&&\tilde n \equiv 
\langle {\widehat \varphi}^{\dagger}{\widehat \varphi} \rangle; 
\hspace{2.7cm}
\tilde m \equiv 
\langle {\widehat \varphi}{\widehat \varphi} \rangle; \\
&&n_T=n_c+\tilde n; 
\hspace{2.4cm}
m_T=m_c+\tilde m. 
\end{eqnarray}
The equation for the classical wave function of the condensate is
closed only when the back-reaction effect due to the fluctuations are
neglected. (This back-reaction is hiding in the parameters $\tilde m$
and $\tilde n$.) This is the approximation contemplated by the
Gross-Pitaevskii equation. In general one will have to solve both
equations simultaneously.

Adopting the Madelung representation for the wave function of the
condensate
\begin{equation}
\psi(t,\x)=\sqrt{n_c(t,\x)} \; \exp[-{\rm i}\theta(t,\x)/\hbar], 
\end{equation}
and defining an irrotational ``velocity field'' by $\v\equiv
{\bnabla\theta}/{m}$, the Gross-Pitaevskii equation can be rewritten
as a continuity equation plus an Euler equation:
\begin{eqnarray}
&&
\frac{\partial}{\partial t}n_c+\bnabla\cdot({n_c \v})=0,
\label{E:continuity}\\
&&
m\frac{\partial}{\partial t}\v+\bnabla\left(\frac{mv^2}{2}+
V_{ext}(t,\x)+\g n_c-
\frac{\hbar^2}{2m}\frac{\nabla^{2}\sqrt{n_c}}{\sqrt{n_c}}
\right)=0.
\label{E:Euler1}
\end{eqnarray}
These equations are completely equivalent to those of an irrotational
and inviscid fluid apart from the existence of the so-called quantum
potential $V_{\rm
  quantum}=-\hbar^2\nabla^{2}\sqrt{n_c}/(2m\sqrt{n_c})$, which has the
dimensions of an energy. Note that
\begin{equation}
n_c \; \nabla_i V_{\rm quantum} 
\equiv 
n_c \; \nabla_i 
\left[
-{\hbar^2\over2m} {\nabla^{2}\sqrt{n_c}\over\sqrt{n_c}} 
\right]
=
\nabla_j \left[ 
-{\hbar^2\over4m} \; n_c \; \nabla_i \nabla_j \ln n_c 
\right],
\end{equation}
which justifies the introduction of the so-called quantum stress tensor
\begin{equation}
\sigma_{ij}^{\rm quantum} =  
-{\hbar^2\over4m} \; n_c \; \nabla_i \nabla_j \ln n_c.
\end{equation}
This tensor has the dimensions of pressure, and may be viewed as an
intrinsically quantum anisotropic pressure contributing to the Euler
equation.  If we write the mass density of the Madelung fluid as $\rho
= m \; n_c$, and use the fact that the flow is irrotational then the
Euler equation takes the form
\begin{equation}
\rho \left[ \frac{\partial} {\partial t}\v+ (\v\cdot\bnabla) \v \right] 
+ \rho \; \bnabla \left[\frac{V_{ext}(t,\x)}{m} \right]
+ \bnabla  \left[{\g \rho^2\over 2 m^2}\right] +
\bnabla \cdot \sigma^{\rm quantum} =0.
\label{E:Euler2}
\end{equation}
Note that the term $V_{ext}/m$ has the dimensions of specific
enthalpy, while $\g \rho^2/(2m)$ represents a bulk pressure.  When the
gradients in the density of the condensate are small one can neglect
the quantum stress term leading to the standard hydrodynamic approximation.
Because the flow is irrotational, the Euler equation is often more
conveniently written in Hamilton--Jacobi form:
\begin{equation}
m \frac{\partial}{\partial t}\theta+ \left( \frac{(\bnabla\theta)^2}{2m}
+V_{ext}(t,\x)+\g n_c-
\frac{\hbar^2}{2m}\frac{\nabla^{2}\sqrt{n_c}}{\sqrt{n_c}}
\right)=0.
\label{E:HJ}
\end{equation}

Apart from the wave function of the condensate itself, we also have to
account for the [typically small] quantum perturbations of the system
(\ref{quantum-field}). These quantum perturbations can be described in
several different ways, here we are interested in the ``quantum
acoustic representation''
\begin{eqnarray}
\widehat \varphi(t,\x)=&&\hspace{-3mm}
e^{-\im \theta/\hbar}
\left({1 \over 2 \sqrt{n_c}} \; \widehat n_1 
- \im {\sqrt{n_c} \over \hbar} \;\widehat \theta_1\right),
\label{representation-change}
\end{eqnarray}
where $\widehat n_1,\widehat\theta_1$ are real quantum fields. 
By using this representation equation (\ref{quantum-field}) can be 
rewritten as 
\begin{eqnarray}
\partial_t \widehat n_1 + 
{1\over m} \bnabla\cdot\left(
\widehat n_1 \; \bnabla \theta + n_c \; \bnabla \widehat \theta_1
\right) = 0, 
\label{pt1} \\
\partial_t \widehat \theta_1  
+  {1\over m} \bnabla \theta \cdot \bnabla \widehat \theta_1 
+ \g(a) \; n_1 - {\hbar^2\over2 m}\; D_2 \widehat n_1 = 0.
\label{pt2}
\end{eqnarray}
Here $D_2$ represents a second-order differential operator obtained
from linearizing the quantum potential. Explicitly:
\begin{eqnarray}
D_2\, \widehat n_1 
&\equiv&
-\half  n_c^{-3/2} \;[\nabla^2 (n_c^{+1/2})]\; \widehat n_1
+\half  n_c^{-1/2} \;\nabla^2 (n_c^{-1/2}\; \widehat n_1).
%\nonumber\\&&
\end{eqnarray}
The equations we have just written can be obtained easily by
linearizing the Gross-Pitaevskii equation around a classical solution:
$n_c \rightarrow n_c + \widehat n_1$, $\phi \rightarrow \phi +
\widehat \phi_1$.  It is important to realize that in those equations
the back-reaction of the quantum fluctuations on the background
solution has been assumed negligible.

We also see in those equations, (\ref{pt1}) and (\ref{pt2}), that time
variations of $V_{ext}$ and time variations of the scattering length
$a$ appear to act in very different ways.  Whereas the external
potential only influences the background equation~(\ref{E:HJ}) [and
hence the acoustic metric in the analog description], the scattering
length directly influences both the perturbation and background
equations.

{From} the previous equations for the linearized perturbations it is
possible to derive a wave equation for $\widehat \theta_{1}$ (or
alternatively, for $\widehat n_{1}$). All we need is to substitute in
equation~(\ref{pt1}) the $\widehat n_{1}$ obtained from
equation~(\ref{pt2}).  This leads to a PDE that is second-order in
time derivatives but infinite order in space derivatives --- to
simplify things we can construct the symmetric $4 \times 4$ matrix
\begin{equation}
f^{\mu\nu}(t,\x) \equiv 
\left[
\matrix{f^{00}&\vdots&f^{0j}\cr
        \cdots\cdots&\cdot&\cdots\cdots\cdots\cdots\cr
        f^{i0}&\vdots&f^{ij}\cr } 
\right].
\label{E:explicit}             
\end{equation}
(Greek indices run from $0$--$3$, while Roman indices run from
$1$--$3$.)  Then, introducing $(3+1)$--dimensional space-time
coordinates --- $x^\mu \equiv (t; x^i)$ --- the wave equation for
$\theta_{1}$ is easily rewritten as
\begin{equation}
\label{E:compact}
\partial_\mu ( f^{\mu\nu} \; \partial_\nu \widehat \theta_1) = 0.
\end{equation}
Where the $f^{\mu\nu}$ are differential operators acting on space only:
\begin{eqnarray}
f^{00} &=& - \left[ \g(a) - {\hbar^2\over 2m}\; D_2 \right]^{-1}
\\
f^{0j} &=& -\left[ \g(a) - {\hbar^2\over 2m}\; D_2 \right]^{-1}\; 
{\nabla^j \theta_0\over m}
\\
f^{i0} &=& - {\nabla^{i} \theta_0\over m} \; 
\left[  \g(a) - {\hbar^2\over2 m}\; D_2 \right]^{-1}
\\
f^{ij} &=& {n_c \; \delta^{ij}\over m} -  
{\nabla^{i} \theta_0\over m} \; 
\left[ \g(a) - {\hbar^2\over2 m}\; D_2 \right]^{-1}\; 
{\nabla^{j} \theta_0\over m}.
\end{eqnarray}
Now, if we make an spectral decomposition of the field $\widehat
\theta_1$ we can see that for wavelengths larger than $\hbar /mc_s$
(this corresponds to the ``healing length'', as we will explain
below), the terms coming from the linearization of the quantum
potential (the $D_2$) can be neglected in the previous expressions, in
which case the $f^{\mu\nu}$ can be approximated by numbers, instead of
differential operators. (This is the heart of the acoustic
approximation.)  Then, by identifying
\begin{equation}
\sqrt{-g} \; g^{\mu\nu}=f^{\mu\nu},
\end{equation} 
the equation for the field $\widehat \theta_1$ becomes that of a
(massless minimally coupled) quantum scalar field over a curved
background
\begin{equation}
\Delta\theta_{1}\equiv\frac{1}{\sqrt{-g}}\;
\partial_{\mu}\left(\sqrt{-g}\; g^{\mu\nu}\; \partial_{\nu}\right)
\widehat\theta_{1}=0,
\end{equation}
with an effective metric of the form
\begin{equation}
g_{\mu\nu}(t,\x) \equiv 
{n_c\over m\; c_s(a,n_c)}
\left[ 
\matrix{-(c_s(a,n_c)^2-v^2)&\vdots& - v_j \cr
        \cdots\cdots\cdots\cdots&\cdot&\cdots\cdots\cr
        -v_i &\vdots&\delta_{ij}\cr } 
\right].
\end{equation}
Here the magnitude $c_s(n_c,a)$ represents the speed of the phonons in the
medium:
\begin{equation}
c_s(a,n_c)^2={\g(a) \; n_c \over m}.
\end{equation}
%

%----------------------------------------------------------------------
\section{Analog models for cosmological spacetimes}
%---------------------------------------------------------------------

To find analog models for cosmological spacetimes we will consider a
generalized GP equation where the external potential and the coupling
constant can both change with time
\begin{equation}
\label{E:tdlg}
 \im \hbar \; \frac{\partial }{\partial t} \psi(t,\x)= \left (
 - {\hbar^2\over2m}\nabla^2 
 + V_{\rm ext}(t,\x)
 + \g(t) \; \left| \psi(t,\x) \right|^2 \right) \psi(t,\x).
\label{eq:GP2}
\end{equation}
The technical steps in the calculation change in a straightforward
manner and lead to a simple time-dependent acoustic
metric
\begin{eqnarray}
ds^2={n_c \over m\;c_s}
\left[-(c_s^2- v^2)\,dt^2-2\vec v \cdot d \vec x \, dt +d\vec x^2 \right]
\label{fluid-geometry}
\end{eqnarray}
it is this time-dependent effective metric that we now
wish to use for simulating a cosmological spacetime and,
subsequently probing cosmological particle production.

%----------------------------------
\subsection{Cosmological analog by explosion}
%----------------------------------

Starting from the geometry (\ref{fluid-geometry}) there are different
ways in which one can reproduce a cosmologically expanding geometry.
Following~\cite{grf,Fischer,Kagan-Castin,Kagan} one can take a radial
profile for the velocity $\vec v=(\dot b/b) \vec r$, with $b$ a scale
factor depending only on $t$, and define a new radial coordinate as
$r_b=r/b$.  In these new coordinates, the metric will be expressed as
\begin{eqnarray}
ds^2={n_c \over m\; c_s}
\left[-c_s^2dt^2+b^2(dr_b^2+r_b^2d\Omega_2^2)\right].
\label{fluid-geometry2}
\end{eqnarray}

Now, one solution for the BEC wave function that reproduces a FRW
universe is this: Introducing a Hubble-like parameter,
\begin{equation}
H_b(t) =  {\dot b(t)\over b(t)}, 
\end{equation}
the equation of continuity can be written as 
\begin{equation}
\dot n_c + 3 H_b(t) \; n_c = 0; 
\qquad \implies \qquad 
n_c(t)= {n_0\over b^3(t)},
\label{eq:rho}
\end{equation}
with $n_0=$constant.  Then, the solution for $\theta$ can be obtained
from (\ref{E:HJ})
\begin{eqnarray}
\theta={\dot b \over 2b}m r^2+\int_0^t dt\; {\g n_0 \over
  b^3(t)} ,
\label{eq:theta-ex}
\end{eqnarray}
and it requires  an external potential of the form
\begin{eqnarray}
V_{\rm ext}(t,r)=-\left({\ddot b \over 2b}+{\dot b^2 \over b^2}\right) m r^2.
\end{eqnarray}
One could certainly construct such a potential in a ``sufficiently
large'' region around $r=0$; this would correspond to a ``sufficiently
large'' part of a homogeneous and flat FRW universe. Outside this
region, the potential will have in practice some confining walls.  The
final effective metric can be written as
\begin{eqnarray}
ds^2=-T^2(t)\; dt^2+a_s^2(t) \; (dr_b^2+r_b^2d\Omega_2^2),
\end{eqnarray}
with
\begin{eqnarray}
a_s(t) \equiv \left({n_0\over m\; \g } \; b \right)^{1/4},
\qquad \hbox{and} \qquad
T(t)\equiv {n_0^3 \over m^3 \; \g^2}
{1 \over a_s^{9}(t)}.
\end{eqnarray}
So finally we end up with a FRW universe whose proper Friedmann time,
$\tau$, is related to the laboratory time, $t$, by $\tau=\int T(t)
dt$.

Although this explosion route seems promising, one should note that
this analog models has substantial drawbacks. In particular it is easy
to see from equation~(\ref{eq:theta-ex}) that one would get for the
condensate a linearly rising velocity field $\v\equiv
{\bnabla\theta}/{m}\propto r$. Hence this particular realization of a
FRW effective geometry is guaranteed to possess an apparent horizon, a
spherical surface in which the speed of the fluid surpasses the speed
of sound. From a dynamical point of view, this might introduce many
practical problems not intrinsically inherent to the type of
geometries we are trying to reproduce. Because of this, we view the
use of an exploding medium as not being the preferred route for
building an analogue for an expanding FRW universe. (For an alternate
view, where the explosion route is the primary focus of attention,
see~\cite{Fischer}.)

%-----------------------------------------------------------
\subsection{Cosmological analog by varying speed of sound}
%-----------------------------------------------------------

Another way to reproduce cosmologically expanding configurations in,
we think, a cleaner fashion is by taking advantage of the possibility
to change the value of the scattering length offered by some
BECs~\cite{Rb1}. Let us note that in reference \cite{Kagan}, the
authors already used the existence of time dependence on the
scattering length, in combination with a time dependent external
potential, to create explosive configuration of the type described in
previous subsection. Here, we are going to use this variability with a
different strategy.

Let us again start from (\ref{fluid-geometry}) but now
with $v=0$ at all times:
\begin{eqnarray}
ds^2=-{n_c c_s \over m} \; dt^2+{n_c \over m\; c_s} \; d\vec
x^2.
\end{eqnarray}
Then, it is not difficult to envisage a situation in which $n_c$ is
constant in a ``sufficiently large'' region (think of a ``sufficiently
large'' close-to-hard-walled box).  Then, the continuity equation is
directly satisfied and the Hamilton--Jacobi equation tells us that
with a fixed external potential the phase function $\theta$ will
depend only on $t$ adapting itself to the changes of $\g(t)$. Changing
the scattering length with time directly causes changes in the value
of the propagation speed $c_s$. (That temporal changes in the velocity
of sound can be interpreted as a cosmological expansion without
invoking any sort of velocity in the medium has already been suggested
in~\cite{Volovik} in the context of superfluid Helium.)  We now define
$\tau= \int [n_c c_s(t)/m]^{1/2} dt$ and write
\begin{eqnarray}
ds^2=-d\tau^2+a_s^2(\tau)\,d\vec x^2, 
\end{eqnarray}
where
\begin{eqnarray}
a_s(\tau)=\left({n_c \over m\; c_s(\tau)}\right)^{1/2}
=\left({n_c  \over m\; \g(a)} \right)^{1/4}
=\left({n_c \over 4 \pi \hbar^2 }\right)^{1/4} {1 \over a^{1/4}}
\equiv {A \over a^{1/4}}.
\label{scale-factor}
\end{eqnarray}
In this model an expansion corresponds to a decrease of the scattering 
length and vice versa. The Hubble function for this spacetime is 
\begin{eqnarray}
H={a_s'\over a_s}=-{1 \over 2} {c_s' \over c_s}
=-{1 \over 2}\left({m \over n_c}\right)^{1/2} \;{\dot c_s \over c_s^{3/2}}.
\end{eqnarray}
(The prime represents derivative with respect Friedmann time, the dot
derivative with respect laboratory time).  This is the model we will
consider in the following discussion.

%----------------------------------------------------------------------
\section{Analog cosmological particle creation}
%---------------------------------------------------------------------

Let us now present a qualitative explanation of how is that we can
closely simulate cosmological particle creation within this model. We
can start with a condensate in a stationary state described, in a
sufficiently large volume, by a constant background density $n_c$ and
a phase function linear in time, $\theta=E_0\, t$. This is a solution
of the GP equation.  For this, one needs to have a potential that
reproduces a large enough close-to-hard-walled box and that satisfies
the condition $E_0=-V_{\rm ext}-\g n_c $.  Apart from this classical
background, there will be some quantum fluctuations over it. At
temperatures much below the critical temperature these quantum
fluctuations are very small and can be described by the Bogoliubov
equations.  (These quantum fluctuations are present even at zero
temperature owing to the so-called quantum depletion phenomenon, see
for example~\cite{Castin}.)  Let us consider that these quantum
fluctuations are in their vacuum state. If one now decreases the value
of the scattering length in a sufficiently slow manner (this issue
will be discussed bellow), all the individual bosons (this is only an
approximate concept in an interacting theory) will be affected in the
same way.  This means that the value of the background magnitudes
---these are the coherent magnitudes--- will be slowly modified.  The
GP equation tells you that the density function $n_c$ will be kept
fixed while the phase function will develop a non-trivial dependence
with time. At the same time the value of the speed of sound will
decrease.

Now, apart from the background configuration, what happens with the
additional quantum fluctuations? The equation satisfied by the quantum
fluctuations is, in the acoustic approximation (that is, for long
wavelengths) that of a massless minimally coupled scalar field over an
expanding background and therefore, it will yield cosmological pair
production of particles.

An interesting point to notice is that varying the external potential
---by this we mean changing with time the value of the external
potential, $V_{\rm ext}(t)$, in the central homogeneous region---
changes the background configuration in the same manner as varying the
scattering length does. However, as the speed of sound does not depend
on the external potential, its variations will not lead to
cosmological particle creation.

Now, what can we say about the observability of the particle
production process?  The standard technique to ``see'' phonons is
tomographic imaging. One opens the trap and looks at the expansion of
the condensate.  Phonons correspond to distributions for the momentum
of the atoms in the trap and different momenta correspond to different
travel distances of the atoms after you switch of the trap. Taking
snapshots of this evolution shows the density contrasts and then the
original momentum distribution.
  
If the wave function $\widehat \Psi$ is split into the condensate wave
function $\psi$ plus quantum excitations $\widehat \varphi$ (in this
situation, we mean atoms) then, the density you observe is
\begin{eqnarray}
 \langle \widehat \Psi^{\dagger} \widehat \Psi \rangle=
 |\psi|^2+\langle \widehat \varphi^{\dagger} \widehat \varphi \rangle,
\end{eqnarray}
as $\langle \widehat \varphi^{\dagger} \rangle =\langle \widehat
\varphi \rangle=0$.  Therefore, the observability of the effect will
depend on the value of the ratio
\begin{eqnarray}
{\langle \widehat \varphi^{\dagger} \widehat \varphi \rangle \over |\psi|^2}=
{1 \over n_c} \langle \widehat \varphi^{\dagger} \widehat \varphi \rangle
\end{eqnarray}
or more simply on the spatially integrated ratio
\begin{eqnarray}
C \equiv {1 \over N_c} 
\int dx^3 \langle \widehat \varphi^{\dagger} \widehat \varphi \rangle.
\label{C-ratio}
\end{eqnarray}
If this quantity is of order unity (say 1/2 or 1/10) then phonons can
probably be ``seen''; if it is 1/100 then seeing phonons is
technologically difficult.  If the particle production process was so
strong that the calculation of $C$ resulted on values close to unity
or higher, this would indicate that the Bogoliubov mean field
approximation would have been violated, and the BEC itself disrupted.

Let us now perform some explicit calculations of the particle
production expected in realistic situations with present-day BECs in
which the scattering length $a$ is changed in time from some initial
value to a different final value. We first have to know how quickly we
can drive these temporal changes while still ensuring that the
different approximations involved in the analysis remain valid.

%---------------------------------------------------------------------
\subsection{Varying $a$, validity of the GP equation}
\label{sec:var-a}
%---------------------------------------------------------------------

The previous analysis shows that in order to consider particle
creation driven by a time-varying scattering length we must be sure to
work in a regime where the background is ``instantaneously'' reacting
to the changes in $a$. Moreover the very derivation of the effective
metric description is based on the GP equation which we then want to
make sure holds at each instant of time.

So we must first determine the upper bound on the rapidity of the
change in the scattering length $a$ which still permits the GP
equation to hold. This will also give an upper bound for the
frequencies of the quasi-particles that might be created (if $\tau$ is
the shortest timescale over which we can drive the system then $\nu
\approx 1/\tau$ is the largest frequency of the quasi-particle we can
create).  The validity of the GP in describing the Bose-Einstein
condensate is related to the validity of several crucial assumptions
which permit us to perform certain approximations on the fundamental
multi-particle Hamiltonian description.  The relevant approximations
are generally stated to be the ``mean-field'' approximation and the
dilute gas approximation.  It is nevertheless important to note that
in a dynamical situation a third approximation, which we can call
``Markovian'' approximation, is implicitly assumed.

Let us review the meaning and implications of these approximations:
The mean field approximation is based on the assumption that most of
the atoms are in the condensate phase and that the influence of the
non condensed fraction can be neglected. This implies that significant
creation of quasi-particles with excessive energies can be dangerous.
In particular from the Bogoliubov dispersion relation~\cite{broken}
\begin{equation}
 \omega=\sqrt{c_s^2 k^2+\left({\hbar \over 2 m}\;k^2\right)^2},
 \label{eq:disprel3}
\end{equation}
we can deduce that excitation of quasi-particles with wavelengths
comparable to the healing length would led to free particle states
(for $k>2\pi/\xi$; $\xi = \hbar/(m\; c_s)$; $\omega \approx \hbar^2
k^2/(2 m)$).  This argument seems to imply that one should require the
typical timescale for the change in $a$ to be no shorter than the
healing time. (Which is the analogue in this situation of the Planck
time in quantum gravity.)

The dilute gas approximation is instead related to the way the
interaction potential is simplified in the GP equation.  This
approximation is valid if $n|a|^3\ll 1$, so we shall have to keep the
amplitude of the changes in $n a^3$ small in order to satisfy this
bound. We wish to emphasise that the dilute gas approximation does not
appear to be a crucial approximation for the analog gravity picture to
hold. As long as the interaction term $\g|\psi(t,\x)|^2$ can be
generalized to be some higher order (but still local) term
$\pi(|\psi|)$ an analog gravity description is not precluded (see, for
example~\cite{bec-cqg}).

Finally the Markovian approximation is related to the fact that in
dynamical situations the two-body time-dependent scattering matrix can
have a complicated form due to the ``memory'' of the system (see e.g.
section IV-A of the paper by K\"ohler and
Burnett~\cite{Kohler-Burnett}).  Basically in these situations the
system is described by a GP-like equation where the interaction term
includes a ``delay term'' described by an integration over different
times. The necessary assumption in order to have a Markovian
description of the dynamics (which together with the two previous
approximations leads to the GP equation) is then that the timescales
on which external parameters are changing are longer that the two-body
collisional duration. That is, longer than the timescale over which a
single interaction happens.  (Reduced to the bare bones, we are asking
that the scattering length does not change significantly during the
period when a pair of atoms are interacting.)

We can estimate the two-body collisional time by a simple calculation.
All we need is the typical size of the region of strong interaction of
two atoms in the condensate and the typical speed with which they
move. The first quantity can be assumed to be of the order of the Van
der Waals scale length: the inter-atomic potential $V(r)$ is
characterized by a short-range region of strong chemical bonding and a
long-range Van der Waals potential,
\begin{equation}
 V \rightarrow -C_6/r^6.
\end{equation}
This leads to a Van der Waals scale
length~\cite{Gribakin93,Weiner99,Williams99,Bolda02},
\begin{equation}
 \lambda_{\rm vdW} = \frac{1}{2} \left(\frac{2\mu C_6}{\hbar^2}\right)^{1/4}.
\end{equation}
This length is basically the size of the region of strong interaction:
for $r \ll \lambda_{\rm vdW}$ the scattering wave function oscillates
rapidly due to the strong interaction potential.  In alkali ground
state interactions, $C_6$ is the same for all hyperfine states of a
given atomic pair; consequently, $\lambda_{\rm vdW}$ is the same for
all collision channels.  For example in the case of Na$_2$, it is
about 2.4 nm. We shall assume here generically $\lambda_{\rm
  vdW}\approx 1$ nm.

Regarding the typical speed of the atoms, this is set by the de
Broglie momentum generated by the trap confinement: $p=h/R$ and
$\bar{v}=p/m$. We shall assume a trap of typical size of ten microns.
We then get
\begin{equation}
t_{\rm int}=\frac{\lambda_{\rm vdW}}{\bar{v}}=
\frac{\lambda_{\rm vdW}\;m\; R}{h}.
\end{equation}
We now confront this quantity with the timescale we have to be faster
than in order to create modes with wavelengths shorter than the
condensate size $R$. This is $t_{\rm size}=R/c_s$. So
\begin{equation}
\frac{t_{\rm int}}{t_{\rm size}}
=\frac{\lambda_{\rm vdW}\;m \;c_s}{h}
=\frac{\lambda_{\rm vdW}}{\xi}
\end{equation}
For typical BEC systems $\xi\approx 1\mu$m--$ 0.1\mu$m (assuming that
the scattering length ranges from one to a hundred nanometers) so
\begin{equation}
\frac{t_{\rm int}}{t_{\rm size}}=\frac{\lambda_{\rm vdW}}{\xi}
\approx 10^{-3}\;\mbox{---}\;10^{-2}.
\end{equation}
Note that $t_{\rm int}$ can be computed to be
\begin{eqnarray}
t_{\rm int}&=&\frac{\lambda_{\rm vdW}}{\bar{v}}
=\frac{\lambda_{\rm vdW}\;m\;R}{h}
\nonumber \\
&=&\frac{10^{-9}\mbox{m}\cdot 85 \cdot 10^9\mbox{ eV}\cdot 10^{-5}\mbox{ m}}
{9\cdot 10^{16} \mbox{ m$^2$/s$^2$}\, 2\pi\cdot 6.5 \cdot 10^{-16}
\mbox{ eV}\cdot\mbox{s}}
\\
 &\approx&2\cdot 10^{-6} \mbox{ s}, \nonumber
\end{eqnarray}
so a microsecond is the shortest timescale allowed for the change in
$a$. Note that this interaction time is shorter than the healing time
$t_{\rm heal}=\xi/c_s\approx 10^{-3}$---$10^{-5} \mbox{ s}$ which
plays the role of the ``Planck time'' in this system. Thus the GP
equation is valid in the entire ``sub-healing'' regime, which is the
primary regime of interest, and continues to hold well into the
``trans-healing'' regime (although the previous comments regarding the
breakdown of the mean-field approximation in this regime remain
valid).

%----------------------------------------------------------------------
\subsection{Analytical calculations: Changing $a$ over a  finite amount 
of time}
%----------------------------------------------------------------------

Now that we have estimated how fast the change in the scattering
length can be driven, we can propose a particular time dependence and
derive an estimate for the relative production of phonons. Particle
production in an expanding universe has been extensively studied in
the framework of semiclassical gravity~(see
e.g.~\cite{Birrell-Davies}). In this regard the scope of this
subsection will be to present an example of these calculations to non
specialists as well as to evaluate the experimental feasibility of an
experimental test.

As a test-bed we shall choose a slightly simplified version of
Parker's model~\cite{Parker}.  The FRW metric with flat spatial
sections can be alternatively written as
\begin{eqnarray}
ds^2=-a_s^6(\eta)\; d\eta^2+a_s^2(\eta)\; d\vec x^2,
\end{eqnarray}
in which we are using for convenience a special type of
pseudotime $\eta=\int(n_c/ma_s^4)dt$ with $a_s=(n_c/mc_s)^{1/2}$.  

The scale factor is independent of $x$ hence the mode decomposition
for the quantum scalar field can be written as
\begin{eqnarray}
\widehat \theta_1(\eta,\x)= 
\int {d\k^3 \over (2 \pi)^{3/2}} 
\left(a_\k e^{i\k \cdot \x}
\psi_k(\eta)
+a_\k^{\dagger} e^{-i\k \cdot \x} \psi_k^*(\eta) \right),
\label{mode-expansion}
\end{eqnarray}
where the $\psi_k(\eta)$ are solutions of the equation
\begin{eqnarray}
{d^2 \psi_k  \over d \eta^2}+ a_s^4(\eta)\,k^2\, \psi_k =0,
\label{mode-equation}
\end{eqnarray}
that satisfy the normalization condition
\begin{eqnarray}
i(\psi_k^* \,\partial_\eta \psi_k- \psi_k \,\partial_\eta \psi_k^*)=1.
\end{eqnarray}
Imagine now that the scale factor $a_s$ undergoes a finite amount of
expansion in a monotonic fashion. This means that the scale factor
passes from and initial value $a_{si}$ at $\eta=-\infty$ to a final
value $a_{sf}$ at $\eta=+\infty$ (or what is the same, that the
scattering length passes from and initial value $a_i$ to a final value
$a_f$). Spacetime can be approximated at early times and at late times
respectively by two different Minkowski spacetimes.  As is standard,
we will assume that the quantum scalar field is initially in the
vacuum state associated to the initial Minkowski spacetime. Then, we
want to calculate the particle content of this state in the final
Minkowski spacetime.

There is a particularly convenient choice of function $a_s(\eta)$ for
which the physics is clear and for which one can exactly solve
equation (\ref{mode-equation}):
\begin{eqnarray}
a_s^4(\eta)={a_{si}^4+a_{sf}^4\over2} 
+  {a_{sf}^4-a_{si}^4\over2} \;
\tanh\left[{\eta\over\eta_0}\right].
\label{eq:prof} 
\end{eqnarray}
(This is a slight simplification of the model considered in
Parker~\cite{Parker}; a variation of this model can be found in the
book by Birrell and Davies~\cite{Birrell-Davies}, see pp. 60 ff.)  If
we now impose as boundary condition for a solution that at $\eta
\simeq -\infty$
\begin{eqnarray}
\psi_k(\eta \rightarrow -\infty)=(2ka_{si}^2)^{-1/2}\exp(-ika_{si}^2\eta),
\label{initial}
\end{eqnarray}
one finds a particular set of exact solutions~\cite{Eckart,Birrell-Davies}
\begin{eqnarray}
\psi^{in}_k(\eta)
&=& {1\over\sqrt{2 k a_{si}^2 }} \exp[-ik(a_{sf}^2+a_{si}^2)\eta/2] 
\; [2\cosh(\eta/\eta_0)]^{-ik\eta_0(a_{sf}^2-a_{si}^2)/2}
\nonumber\\
&&
\times \; _2F_1\left(1+ik\eta_0[a_{sf}^2-a_{si}^2]/2,+ik\eta_0[a_{sf}^2-a_{si}^2]/2; 
1-ik\eta_0 a_{si}^2;  {1\over2}[1+\tanh(\eta/\eta_0)] \right).
\end{eqnarray}
These solutions correspond to positive {\em in-going} modes.
Similarly, the precise form of the exact solutions that at $\eta
\simeq +\infty$ admit the asymptotic form
\begin{eqnarray}
\psi_k(\eta \rightarrow +\infty)=(2ka_{sf}^2)^{-1/2}\exp(-ika_{sf}^2\eta),
\label{final}
\end{eqnarray}
(the positive {\em out-going} modes) is
\begin{eqnarray}
\psi^{out}_k(\eta) &=& {1\over\sqrt{2 k a_{sf}^2 }}
\exp[ik(a_{sf}^2+a_{si}^2)\eta/2] \;
    [2\cosh(\eta/\eta_0)]^{ik\eta_0(a_{sf}^2-a_{si}^2)/2}
    \nonumber\\ && \times \;
    _2F_1\left(1-ik\eta_0[a_{sf}^2-a_{si}^2]/2,-ik\eta_0[a_{sf}^2-a_{si}^2]/2;
    1-ik\eta_0 a_{si}^2; {1\over2} [1-\tanh(\eta/\eta_0)]
    \right).
\end{eqnarray}
The Bogoliubov coefficients relating the early time ({\em in-going}) 
and late time ({\em out-going}) bases are then defined as
\begin{eqnarray}
\psi_k^{in}({\eta \rightarrow \infty})={1\over\sqrt{2ka_{sf}^2}} \; 
[\alpha_k \exp(-ika_{sf}^2\eta)+\beta_k \exp(ika_{sf}^2\eta)].
\label{bogo}
\end{eqnarray}
{From} this expression, and the exact solutions above, one can show
\cite{Parker,Birrell-Davies,Eckart,sono-I} that
\begin{eqnarray}
\alpha_k &=& {2 a_{sf} a_{si}\over a_{sf}^2+a_{si}^2   } \;
{\Gamma(-ik\eta_0 a_{si}^2) \; \Gamma(-ik\eta_0 a_{sf}^2) 
\over
\Gamma(-ik\eta_0[a_{sf}^2+a_{si}^2]/2) \; 
\Gamma(-ik\eta_0[a_{sf}^2+a_{si}^2]/2)
},
\\
\beta_k &=& -{2 a_{sf} a_{si}\over a_{sf}^2-a_{si}^2} \;
{
\Gamma(-ik\eta_0 a_{si}^2) \; \Gamma(+ik\eta_0 a_{sf}^2) 
\over
\Gamma(+ik\eta_0[a_{sf}^2-a_{si}^2]/2) \; 
\Gamma(+ik\eta_0[a_{sf}^2-a_{si}^2]/2)
},
\end{eqnarray}
and that
\begin{eqnarray}
|\alpha_k|^2 &=& {
\sinh^2(\pi k \eta_0 [a_{sf}^2+a_{si}^2]/2)
\over
\sinh(\pi k \eta_0 a_{si}^2)\sinh(\pi k \eta_0 a_{sf}^2)
},
\\
|\beta_k|^2 &=& {
\sinh^2(\pi k \eta_0 [a_{sf}^2-a_{si}^2]/2)
\over
\sinh(\pi k \eta_0 a_{si}^2)\sinh(\pi k \eta_0 a_{sf}^2)
}.
\end{eqnarray}
These expressions are related to the Bogoliubov coefficients by
[\emph{cf} Birrell-Davies equation (3.93)]
\begin{equation}
\alpha(\vec k_{in},\vec k_{out}) = 
\alpha_k \; \delta^3( \vec k_{in} - \vec k_{out}),
\end{equation}
\begin{equation}
\beta(\vec k_{in},\vec k_{out}) = 
\beta_k \; \delta^3( \vec k_{in} + \vec k_{out} ).
\end{equation}
The spectrum of particles in the final state is then
\begin{equation}
{\d N\over \d^3 k_{out} } = 
\int \left|\beta(\vec k_{in},\vec k_{out})\right|^2 \d^3 k_{in},
\end{equation}
which gives
\begin{equation}
{\d N\over \d^3 k_{out} } = \int |\beta_k|^2  \; \delta^3(\,\vec 0) \; 
\delta^3( \vec k_{in} + \vec k_{out}) \; \d^3 k_{in}
= |\beta_k|^2  \; \delta^3(\,\vec 0).
\end{equation}
We now use the standard scattering theory result that a momentum-space
delta function evaluated at zero is proportional to the volume of the
``universe'', in this case the volume of the BEC,
\begin{equation}
\delta(\,\vec 0) \to {V\over(2\pi)^3}
\end{equation}
to see that
\begin{equation}
{\d N\over \d^3 k_{out} } = {V\over(2\pi)^3} |\beta_k|^2.
\end{equation}
Equivalently
\begin{equation}
{\d N\over\d k_{out}} = {V\over2\pi^2} |\beta_k|^2 \;
k_{out}^2,
\end{equation}
and the total number of emitted phonons is
\begin{equation}
N =  {V\over2\pi^2} \int_0^\infty  |\beta_k|^2 \; k^2 \d k.
\end{equation}
As a practical matter the integral will always be cut off at high
momentum --- most typically by the inverse timescale $\eta_0^{-1}$
over which the propagation speed changes, but if nothing else the
integral cannot be trusted for momenta higher than that associated
with the healing length $k_{\rm heal}=2\pi/\xi$ for the reasons
previously discussed (see Sec.~\ref{sec:var-a}).

In order to gain a better understanding of the particle creation just
described, it may be useful to study separately the two opposite
regimes characterizing this phenomenon. In fact, for a given timescale
of change, $\eta_0$, driving the particle creation, one has a
simplified description of the particle production when considering the
case of modes with frequencies much smaller than $1/\eta_0$ (sudden
approximation) or much larger than that (adiabatic approximation).
After a brief discussion of these regimes we shall deal with the full
intermediate case.

%----------------------------------------------------------------------
\subsubsection{Sudden approximation}
%----------------------------------------------------------------------

A particularly useful approximation is to take the ``sudden limit''.
Mathematically this consists of taking a step function for the scale
factor
\begin{eqnarray}
a_s^4(\eta)=a_{si}^4+\Theta(\eta) \; [a_{sf}^4-a_{si}^4].
\end{eqnarray}
Physically this means that one is considering that the change in
$a_s(\eta)$ is driven more rapidly than the frequency band one is
interested in. In this case, this means that the change is so fast
that the entire acoustic regime is excited (the time rate is
trans-healing), but sufficiently slowly that the GP equation still
continues to hold (the time rate is still sub-interaction). However we
shall still have to put in ``by hand'' a high momentum cut off, given
by the healing length $k_\mathrm{max}= k_{\rm heal}= 2\pi/\xi$,
because beyond this point we cannot trust the dispersion relation to
remain on the acoustic branch implicit in our calculation.

The relevant calculation can be preformed by simply considering the
mathematical $\eta_0\to 0$ limit in Parker's model. Indeed the
$\alpha$ and $\beta$ coefficients are now momentum independent and
\begin{equation}
|\alpha| = { a_{sf}^2+a_{si}^2\over 2 a_{sf} a_{si} },
\end{equation}
\begin{equation}
|\beta| = { |a_{sf}^2-a_{si}^2|\over 2 a_{sf} a_{si} }.
\end{equation}
As should be expected, particle production in this sudden limit
depends only on the change in the scale factor. The particle
production spectrum is now flat (more precisely, phase space limited)
all the way up to the healing frequency.  A rigorous result is that
for \emph{any} monotonic change in $a_s(\eta)$ from $a_{si}$ to
$a_{sf}$ the magnitudes of the $\alpha$ and $\beta$ coefficients are
less than or equal to those calculated for the sudden
approximation~\cite{bounds} --- consequently the sudden approximation
provides an absolute upper bound on particle production.
  
The number of phonons produced is
\begin{equation}
N = {V\over6\pi^2}  |\beta|^2 k_\mathrm{max}^3 
= {V\over6\pi^2}  |\beta|^2 (2\pi/\xi)^3 
= {4\pi\over3} {V\over \xi^3} |\beta|^2.
\end{equation}
That is
\begin{equation}
N = {4\pi\over3} {V\over \xi^3} { |a_{sf}^2-a_{si}^2|^2\over 4 a_{sf}^2 a_{si}^2 }.
\label{N-sudden}
\end{equation}
The good news for current purposes is that this scales as $(R/\xi)^3$.
Now the trap size $R$ is of order 10 microns, while the healing length
is in the range from 1 to 0.1 micrometers; thus $ (R/\xi)^3 \approx
10^3$ to $10^{6}$.  A prefactor this big is desirable in terms of
producing an observable effect.  As for $a$ (scattering length), this
can range from 100~nm to 1~nm, so the ratio $a_i/a_f$ is up to
$\approx 100$.  Since $a_s^2$ (scale factor) is $\propto a^{-1/2}$ we
have $a_{sf}^2/a_{si}^2$ up to $\approx 10$. Therefore
\begin{equation}
|\beta|^2 = { |a_{sf}^2-a_{si}^2|^2\over 4 a_{sf}^2 a_{si}^2 } \lesssim 2
\end{equation}
which is of order one --- so there is no enormous suppression coming
from the Bogoliubov factor.  All in all, we estimate that $N\approx
10^4$ to $10^{7}$ phonons can be produced in the sudden approximation.

%----------------------------------------------------------------------
\subsubsection{Adiabatic approximation}
%----------------------------------------------------------------------

In contrast, when the parameter $\eta_0\, k\, a_{sf}^2$ is large
compared with unity (that is, for large enough momenta), and provided
$a_{sf}\gg a_{si}$, we have
\begin{eqnarray}
|\beta_k|^2 \simeq {1\over\exp(2\pi\eta_0 k a_{si}^2)-1},
\end{eqnarray}
so that the spectrum of phonons in the final state is
\begin{equation}
{\d N\over \d V \d^3 k} \simeq {1\over(2\pi)^3}  
{1\over\exp(2\pi\eta_0 k a_{si}^2)-1}.
\label{eq:spect-ad}
\end{equation}
This is a correctly normalized Planckian [black body] distribution.
We can associate a temperature $T$ to the final phonon content
produced by the expansion. Before doing that, let us write down some
useful relations between the frequencies associated with the
pseudotime $\eta$ and with the laboratory time $t$. Asymptotically
(either in the infinite past or in the infinite future) the relation
between the times $t$ and $\eta$ is given by $t\sim
a^{2}_{s}\eta/c_s+{\rm const}\sim (ma_s^4/n_c)\eta+{\rm const}$ so
that
\begin{eqnarray}
\omega^{t}_{i}&=&\frac{n_c}{ma_{si}^4}\;\omega^{\eta}_{i} 
=\frac{c_{si}}{a_{si}^{2}}\;\omega^{\eta}_{i}
= c_{si}\;k; 
\\
\omega^{t}_{f}&=&\frac{n_c}{ma_{sf}^4}\;\omega^{\eta}_{f} 
=\frac{c_{sf}}{a_{sf}^{2}}\; \omega^{\eta}_{f}
= c_{sf}\;k.
\label{omega-t-eta}
\end{eqnarray}
Here, the upper indices indicate with respect to which time a
frequency is defined, and the sub-scripts $i$ and $f$ identify whether
a particular magnitude is evaluated in the initial or in the final
configuration. Now it is easy to see that the laboratory temperature
that we would associate to the final configuration in the adiabatic
approximation is
\begin{eqnarray}
T ={1\over 2\pi k_B} \hbar \; c_{sf} \; \eta_0^{-1} a_{si}^{-2}.
\end{eqnarray}

In order to estimate this temperature we need to convert the time
scale over which the scattering length changes, $\eta_{0}$ from the
pseudo time to the laboratory time. To do this we define
\begin{equation}
t_{0}\equiv\eta_{0}\left.\frac{dt}{d\eta}\right|_{\eta=0}.
\label{t-eta-rel}
\end{equation}
For the particular profile, equation~(\ref{eq:prof}), that associated
with this model equation~(\ref{t-eta-rel}) evaluates to
\begin{equation}
t_{0}=\frac{m}{n_c}\left(\frac{a^{4}_{si}+a^{4}_{sf}}{2}\right)\eta_0
=\left(\frac{a^{4}_{si}+a^{4}_{sf}}{2 a^{2}_{sf} c_{sf}}\right)\eta_0
\label{t-eta-rel-2}
\end{equation}
If we use this relation between $t_0$ and $\eta_0$, the temperature
would be
\begin{eqnarray}
T= {1\over 4\pi k_B} \; \hbar \; 
\frac{1}{t_{0}} \;
\frac{a^4_{si}+a^{4}_{sf}} {a^{2}_{sf}a^{2}_{si}}.
\label{temperature}
\end{eqnarray}
In this way we check the intuitive idea that in the expansion process
one would create phonons with frequencies inversely related with the
temporal scale of change of the configuration in laboratory. The
number of phonons in the final state is
\begin{equation}
N \approx 
{V \zeta(3)\over\pi^2} \; {1\over(2\pi \eta_0 a_{si}^2)^3} 
= {V \zeta(3)\over 8 \pi^5 \eta_0^3 a_{si}^6}
={\zeta(3) \over 64 \pi^5}{V \over (c_{sf}t_0)^3}
\left(\frac{a^4_{si}+a^{4}_{sf}}{a_{sf}^{2}a_{si}^{2}}\right)^{3}
\end{equation}

Contrary to what happens in the sudden approximation, the adiabatic
approximation is taking into account the rapidity with which the
configuration changes.  Thus, the total number of phonons calculated
by trusting the adiabatic approximation throughout the whole range of
frequencies is suppressed with respect to the sudden approximation
calculation by a factor of $0.1$ times $(t_{\rm heal}/t_0)^3 = 0.1
\times [({\rm healing~time})/({\rm evolution~time})]^3$ --- this
factor consisting of a dimensionless number coming from the detailed
expression for the integral, times the cubed ratio of the healing time
in the condensate to the time over which the scattering length is
forced to change. With $t_0\approx 10\; t_{\rm heal}$ we still get
somewhere between $100$ and $10^5$ phonons.  Once $t_0 \approx 100 \;
t_{\rm heal}$ this is reduced to somewhere between $10^{-1}$ and $100$
phonons; but in this case we will also run into problems from finite
volume effects --- $t_0$ is then comparable to the sound crossing time
for the condensate and the momentum space delta functions appearing
above are smeared out due to the finite volume of the condensate.
(This point is carefully addressed in a rather different physical
context in~\cite{finite-volume}, though many of the mathematical
manipulations appearing therein are very similar to the present
situation.)

Note that as $t_0\to t_{\rm heal}$ the adiabatic approximation
calculation still results in one order of magnitude less than the
equivalent calculation with the sudden approximation.  Moreover, in
this case, the whole range of observable phonons (with frequencies
between the healing frequency and the trap frequency) is beyond the
strict range of applicability of the adiabatic approximation (remember
that $\eta_0k a_{sf}^2 \gg 1$). Therefore, to be more precise one will
have to make an intermediate analysis, in between the sudden and the
adiabatic regimes.

Let us estimate the value of the temperatures associated with the
adiabatic approximation for temporal scales of change within the
observable window. From equation (\ref{temperature}) we can see that
this temperature will be $T \simeq 10^{-11} {\rm K \cdot s} \cdot
t_0^{-1}$, and so, for $t_0$ between $10^{-2}$~s (associated with the
condensate size) and $10^{-5}$~s (the shortest timescale compatible
with the acoustic approximation associated with the initial
configuration) will range from $1$~nK to $1000$~nK. From this estimate
we can already see that, by modifying the scattering length on time
scales close to the healing time, one could produce a bath of (almost)
thermal phonons so energetic that even the mean field approximation
might break down (causing the complete disruption of the condensate).

When the peak frequency tends toward the healing frequency we see that
the low frequency part of the observable spectrum will develop
important departures from thermality. We shall now turn to the general
intermediate case and describe the full spectrum of phonons created in
our toy model. Then, we will discuss the observability of the
cosmological particle creation effect in terms of the ratio $C$
defined in equation (\ref{C-ratio}).

%----------------------------------------------------------------------
\subsubsection{Intermediate regime}
%----------------------------------------------------------------------

Let us now consider an intermediate regime: Look at large momenta and
ignore for the time being any cutoff arising from the interaction
timescale or the healing length, then
\begin{equation}
|\beta|^2 \to \exp[-4\pi\eta_0\; k\; a_{si}^2] 
\qquad \hbox{as} \qquad
k \to \infty,
\end{equation}
so the particle spectrum is always exponentially suppressed at
sufficiently high momenta. If we pick $\eta_0$ to be longer than a
healing time (so that we cannot use the sudden approximation) but
still sufficiently short that we cannot use the adiabatic
approximation, then we will need to retain the full form of
$|\beta|^2$. The resulting spectrum of phonons is
\begin{equation}
{\d N\over \d V \d^3 k} = {1\over(2\pi)^3} 
{\sinh^2[\pi \eta_0 k (a_{sf}^2-a_{si}^2)/2]
\over
\sinh[\pi \eta_0 k a_{sf}^2] \; \sinh[\pi \eta_0 k
  a_{si}^2]  },
\label{eq:num-spec}
\end{equation}
and the total number of phonons produced is
\begin{equation}
N =  {V\over2\pi^2} \int_0^\infty k^2 \;
{\sinh^2[\pi \eta_0 k (a_{sf}^2-a_{si}^2)/2]
\over
\sinh[\pi \eta_0 k a_{sf}^2] \; \sinh[\pi \eta_0 k a_{si}^2]  } \;\d k,
\end{equation}
We can now consider the actual spectrum described by
equation~(\ref{eq:num-spec}) by choosing plausible values for an
experiment. However in order to get the spectrum that might be
observed one has to convert the relevant quantities in
equation~(\ref{eq:num-spec}) to the laboratory counterparts.

Using the expressions in equation (\ref{omega-t-eta}),
and the relation between $t_0$ and $\eta_0$ given in
equation~(\ref{t-eta-rel-2}), and alternatively rewritten
here as
\begin{equation}
t_{0}
=\left(\frac{a^{4}_{si}+a^{4}_{sf}}{2 a^{2}_{sf} c_{sf}}\right)\eta_0
=\left(\frac{a^{4}_{si}+a^{4}_{sf}}{2 a^{2}_{si} c_{si}}\right)\eta_0,
\end{equation}
the number spectrum can be written as (we are reintroducing here the
explicit {\em in} and {\em out} indices)
\begin{equation}
 {\d N\over\d^3 \vec{k}_{out}} =
 {V\over(2\pi)^3}\int\frac {\sinh^2\left[\pi\,t_0
     \frac{\textstyle
       a^{4}_{sf}\omega^t_{out}-a^{4}_{si}\omega^t_{in}}
          {\textstyle a^{4}_{si}+a^{4}_{sf}}\right]}
 {\sinh\left[2\pi \frac{\textstyle a^{4}_{sf}}
     {\textstyle
       a^{4}_{si}+a^{4}_{sf}}\omega^t_{out}t_0\right]\,\sinh\left[2\pi
     \frac{\textstyle a^{4}_{si}} {\textstyle
       a^{4}_{si}+a^{4}_{sf}}\omega^t_{in}t_0\right]}
 \delta\left(\vec{k}_{in}+\vec{k}_{out}\right)\d^3\vec{k}_{in},
\label{eq:num-spec2}
\end{equation}
which implies
\begin{equation}
 {\d N\over\d \omega^t_{out}} = {V\over{2\pi^2}}
\frac
 {\sinh^2\left[\pi
\frac{ \textstyle \left(a^{2}_{sf}-a^{2}_{si}\right)a^{2}_{sf}}
 {\textstyle a^{4}_{si}+a^{4}_{sf}}
\omega^t_{out}t_0
 \right]} 
{
\sinh\left[2\pi \frac{\textstyle a^{4}_{sf}} {\textstyle
    a^{4}_{si}+a^{4}_{sf}}
\omega^t_{out} t_0\right]\,
\sinh\left[2\pi \frac{\textstyle a^{2}_{si}a^{2}_{sf}}
  {\textstyle a^{4}_{si}+a^{4}_{sf}}\omega^t_{out}t_0\right]}
%\left(\frac {ma_{sf}^{2}} {n_{c}} \right)^{3} 
\frac{(\omega^{t }_{out})^2}{c_{sf}^3}.
\label{eq:num-spec3}
\end{equation}
Regarding the relative range of the scale factor $a_s$, we have
already seen that it can be deduced from the experimentally plausible
range for the scattering length and we shall take
$a_{sf}^2/a_{si}^2\approx 10$.

For the final value of the speed of sound $c_{sf}$ we shall take
$c_{sf}\sim 10$~mm/s. In fact also the range of the speed of sound can
be determined from the scattering length. This can be reasonably be
varied in the proximity of a Feshbach resonance from 1~nm to 100~nm.
For an experiment with heavy alkali atoms (e.g.  rubidium) the speed
of sound will then typically range from few mm/s to 10~mm/s.

%%%==========================================================
\begin{figure}[htb]
\vbox{\hfil
\scalebox{0.50}{\includegraphics{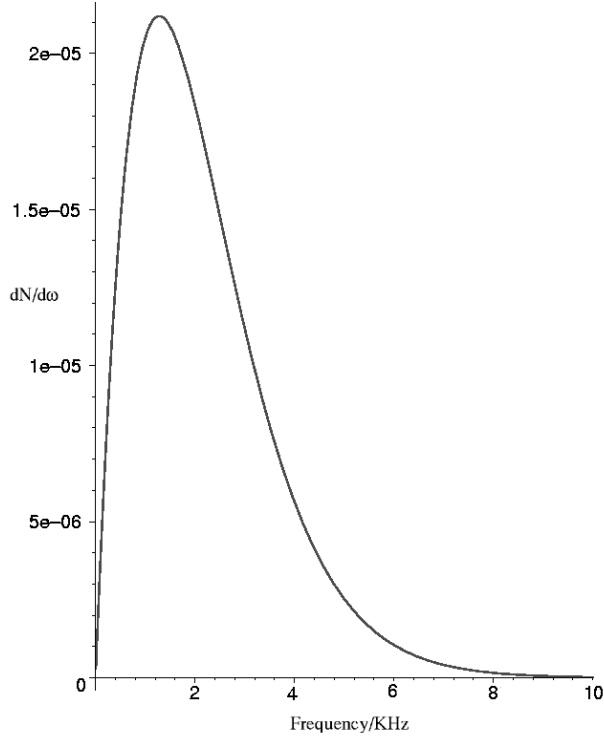}}\hfil}
\caption{
Number spectrum, equation~(\ref{eq:num-spec3}), as
phonons per 1000 cubic microns per unit out frequency.
We have set $c_{f}=10$mm/s and $a_{sf}^2/a_{si}^2\sim 10$.  
The typical timescale $t_{0}$ is conservatively set equal to $10^{-3}$s.}
\label{fig:spectrum} 
\end{figure}
%%%==========================================================

The total number of phonons emitted is given by
\begin{equation}
N =  {1 \over 16 \pi^5} 
{V \over (c_{sf}t_0)^3} 
\left({a_{sf}^4+a_{si}^4 \over a_{sf}^2 a_{si}^2}\right)^3
\; \int_0^\infty x^2 \;
{\sinh^2[x(a_{sf}^2/a_{si}^2-1)/2]
\over
\sinh[x (a_{sf}^2/a_{si}^2)] \; \sinh[x]  } \;\d x =
{1 \over 16 \pi^5} 
{V \over (c_{sf}t_0)^3} 
\left({a_{sf}^4+a_{si}^4 \over a_{sf}^2 a_{si}^2}\right)^3
F(a_{sf}^2/a_{si}^2),
\end{equation}
so qualitatively we have the same behaviour as in the adiabatic
approximation, modulated by a dimensionless function of the ratio
$a_{sf}/a_{si}$. (This expression will remain valid as long as
$\eta_0$ is longer than the healing time; at which stage one should
switch over to the sudden approximation.) Using hyperbolic trig
identities the previous integral can be re-written as
\begin{equation}
F(z) = {1\over2} \int_0^\infty x^2 
\left\{ \coth(zx)\; \coth(x) -1 - {1\over\sinh(zx)\;\sinh(x)} \right\} 
\; \d x
\end{equation}
with
\begin{equation}
F(1)=0; 
\qquad
F(\infty) = {\zeta(3)\over4} \approx 0.3005;
\qquad \hbox{and} \qquad
F(1/z) = z^3 F(z).
\end{equation}
For $z>1$ the function $F(z)$ quickly and smoothly approaches its
asymptotic value.

\begin{figure}[htb]
\vbox{\hfil
\scalebox{0.50}{\includegraphics{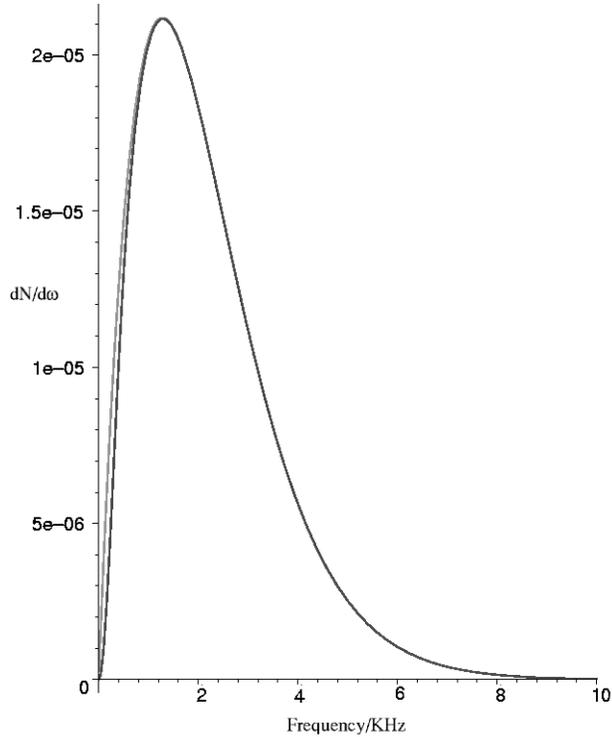}}\hfil}
\caption{
  Comparison of the exact number spectrum, equation~(\ref{eq:num-spec3})
  darker line in the graph, with the one obtained in the adiabatic
  approximation, equation~(\ref{eq:num-spec-ad}). Both the spectra are
  shown as phonons per 1000 cubic microns per unit out frequency.  In
  both the cases we have set $c_{f}=10$mm/s and $a_{sf}^2/a_{si}^2\sim
  10$.  The typical timescale $t_{0}$ is conservatively set equal to
  $10^{-3}$s.}
\label{fig:spectrum2} 
\end{figure}

In closing our analysis of the intermediate regime we want now to
compare the exact spectrum equation~(\ref{eq:num-spec3}) with the
spectrum obtained within the adiabatic approximation.  Equation
(\ref{eq:spect-ad}) can be easily rewritten in the laboratory
variables as
\begin{equation}
 {\d N\over\d \omega^t_{out}} = {V\over{2\pi^2}}
\frac{1} 
{
\exp\left[4\pi \frac{\textstyle a^{2}_{sf}a^{2}_{si}
} {\textstyle
    a^{4}_{si}+a^{4}_{sf}}
\omega^t_{out} t_0\right]-1
}
\frac{(\omega^{t }_{out})^2}{c_{sf}^3}.
\label{eq:num-spec-ad}
\end{equation}
The combined plot of (\ref{eq:num-spec3}) and (\ref{eq:num-spec-ad})
is shown in figure~\ref{fig:spectrum2}.

%%%==========================================================
%%%==========================================================

It is evident that apart from a minor discrepancy at lower frequencies
the two plots basically coincide from the peak frequency
(approximately 2 kHz) on. This is not so surprising given that the
adiabatic approximation implies $\eta_0 k a_{sf}^{2}\gg 1$ which in
laboratory variables corresponds to
\begin{equation}
2 \frac{\textstyle a^{2}_{si}a^{2}_{sf}}
  {\textstyle a^{4}_{si}+a^{4}_{sf}}\omega^t_{out}t_0\gg 1.
\end{equation}
One can easily check that this inequality starts to be
satisfied for frequencies of the order of a few kHz (and
holds for any larger frequency).

%----------------------------------------------------------------------
\subsection{Observability of the effect}
%---------------------------------------------------------------------

In this subsection we will calculate the relevant ratio $C$ defined in
equation (\ref{C-ratio}).  We want $C\geq 1/100$ to make the phonons
easily detectable.  By using equation (\ref{representation-change}) we can
deduce that
\begin{eqnarray}
C={1 \over N_c} 
\int dx^3 \left({1 \over 4 n_c} \langle \widehat n_1 \widehat n_1\rangle
+{n_c \over \hbar^2} \langle \widehat \theta_1 \widehat \theta_1\rangle
-i{1 \over 2\hbar}\langle [\widehat n_1,\widehat \theta_1] \rangle
\right).
\end{eqnarray}
In the acoustic regime and for the particular configurations we are
looking at we know that $\g \widehat n_1=-\partial_t \widehat
\theta_1$.  Considering now that $[\widehat n_1(x,t),\widehat
\theta_1(y,t)]=i \hbar \; \delta^3(x-y)$ or $[\widehat
\theta_1(x,t),\partial_t \widehat \theta_1(y,t)]=i \hbar \; \g \;
\delta^3(x-y)$ --- we can deduce this starting from the fundamental
commutation relation $[\widehat \Psi^\dagger (x,t),\widehat
\Psi(y,t)]=\delta^3(x-y)$)--- we can arrive at
\begin{eqnarray}
C={1 \over N_c} 
\int dx^3 \left(
{1 \over 4 n_c \g^2} 
\langle \partial_t \widehat \theta_1 \partial_t \widehat \theta_1 \rangle
+{n_c \over \hbar^2} \langle \widehat \theta_1 \widehat \theta_1 \rangle
+{1 \over 2} \delta(0)
\right).
\end{eqnarray}
To calculate this average, we can expand the real field operator
$\widehat \theta_1$ in terms of the creation and annihilation
operators associated with the final configuration
\begin{eqnarray}
\widehat \theta_1(y,t)=\int dk^3 
{(\hbar \g)^{1/2} \over (2\pi)^{3/2}(2\omega(k))^{1/2} }
\; a_k \; e^{-i\omega t+i\k \cdot \x} + {\rm H.C.}
\end{eqnarray}
and consider the appropriate particle content for the quantum state.
If $N_k$ is the average number of particles with momentum $k$ in the
quantum state, the previous magnitude can be expressed as
\begin{eqnarray}
C=&&\hspace{-3mm}{1 \over 2 N_n \; (n_c \g)} \int \hbar \omega(k) \; N_k dk^3+
2 {(n_c \g) \over N_c} \int  {1 \over \hbar \omega(k)} N_k dk^3 \\
=&&\hspace{-3mm}
{1 \over 2 N_c\; (mc_s^2)} \int \hbar \omega(k) \; N_k dk^3+
2 {mc_s^2 \over N_c} \int  {1 \over \hbar \omega(k)} N_k dk^3.
\end{eqnarray}
To reach this expression we have to substract the vacuum contribution.
If we consider now the sudden approximation to calculate this rate (an
upper limit to what one could get) we obtain
\begin{eqnarray}
C= 
\left\{ {3 \over 8}{\hbar \omega_{\rm heal} \over mc_s^2} +
3 {mc_s^2 \over \hbar \omega_{\rm heal}} \right\} \; {N \over N_c},
\end{eqnarray}
with $N$ the one in equation (\ref{N-sudden}).  Now $\hbar \omega_{\rm
  heal} \simeq 10^{-10 \pm 1}$~eV. Instead $mc_s^2 \simeq {\rm atomic
  \; number} \times 10^{-13 \pm 1}$~eV $\simeq 10^{-11 \pm 1}$~eV.
The factor $\hbar \omega_{\rm heal}/mc^2 \simeq 10$, so we have that
the relevant number $C \simeq 4 N/N_c$. But $N_c \simeq 10^{6}$ and $N
\simeq 10^{4}-10^{7}$. This gives $C \simeq 4 \times 10^{-2}- 40 $.
This number is based on the sudden approximation and will be smaller
in a more realistic calculation. However, remembering the discussion
on the adiabatic approximation we know that, for temporal scales of
change $t_0$ of the order of the healing time, the actual coefficient
$C$ cannot be smaller than $0.1$ times the previous estimate, i.e.  $C
\simeq 4 \times 10^{-3}- 4 $.  Therefore, by implementing in
laboratory an expansive process with $t_0$ in an intermediate regime,
in between the healing times $t_{\rm heal}=\xi /c_s$ associated with
the initial configuration ($t_0 \simeq 10^{-5}$ seconds) and the final
configuration ($t_0 \simeq 10^{-3}$ seconds), one should be able to
observe the effect.

%----------------------------------------------------------------------
\section{Summary and Discussion}
%---------------------------------------------------------------------

In this work we have discussed the possibilities that BECs offer for
simulating Lorentzian geometries of the cosmological type in the
laboratory. There are two inequivalent paths one can follow in this
task.  The first one is based on provoking an expansive explosion in
the condensate by changing with time the characteristic frequency of
an isotropic and harmonic confining potential. This option implies
that the velocity profile of the condensate acquires arbitrarily high
values at large distances from the center. So there is always a sphere
at which the velocity of the expanding BEC would surpass its sound
velocity: A sonic horizon would be formed.  In practice, due to the
fact that physically realizable BECs are finite systems, one can only
reproduce on them a portion of an expanding universe. Therefore one
might argue that the sonic horizon would be formed outside the BEC.
However, the velocity of sound in BECs is so small (a few mm/s) that
in practical situations the sonic horizon will be formed well inside
the system. Now the existence of sonic horizons is certainly
interesting in its own right, but is not inherent to the simulation of
cosmological spacetimes. Moreover, the plausible dynamical
instabilities associated with their formation could mask the
observation of purely cosmological effects.
  
The alternative path to simulating a cosmological geometry that we
have pursued in this article is to take advantage of the possibility
of varying the scattering length or, what is the same, the interaction
strength between the atoms in the condensate. In this case, what we
need is a confining potential with a sufficiently large almost flat
minimum in which a portion of the condensate stays at rest. In this
configuration, there is no formation of sonic horizons and thus we
think it is (both conceptually and technically) a much cleaner path to
follow.
  
The description of the condensation phenomenon naturally involves the
separation of the system into a ``classical'' wave function (the
condensate part) and quantum fluctuations. In the acoustic
approximation we can think of these quantum fluctuations as phonons
over a classical background geometry, in this case, the analogue of a
cosmological spacetime.  Therefore, we can use the tuning of the
scattering length to simulate not only a classical cosmologically
expanding universe, but the quantum phenomenon of cosmological
particle creation.  We have analyzed this well known process by using
a minor variant of Parker's model for a finite amount of
expansion~\cite{Parker}. Then, by working with numerical estimates
appropriate to currently accessible BECs in dilute gases, we have
presented an analysis of the feasibility of observing the effect in
real experiments.
  
We have seen that there is a more than plausible window for the
observability of the effect with current technology.  In current BECs
the scattering length can easily be varied from a 100~nm to 1~nm.
This produces an expansion in the geometrical scale factor of about
three times. The temporal scale of change of the scattering length
cannot be arbitrarily short. It has to be slower than the time scale
in which the interaction between two atoms proceeds. We have
calculated this time scale to be of the order of microseconds.
However, we have also seen that, by driving the previous finite amount
of expansion in a temporal scale of change of about fractions of
millisecond, one could start to detect the presence of cosmological
particle creation.  {From} here one could shorten the time scale down
to tens of microsecond progressively amplifying the expected effect.
By the time one reaches tens of microseconds the effect would have
been amplified by a factor of a hundred, (with timescale still above
the interaction time), opening even the possibility of totally
disrupting the condensate.
  
The relevant temporal scales of change for the effect to be observable
are of the order of the healing time in the condensate.  Therefore we
expect that apart from the phonon spectrum calculated by neglecting
the modified dispersion relations at high energies, there will be also
some production of quasiparticles.  To observe the purely cosmological
effect one would have to keep this quasiparticle production under a
certain level; thus, the temporal scale of change should not be driven
significantly beyond the healing time.

In our analysis we have neglected finite volume effects.  However we
shall now show that these effects are insignificant for the typical
condensate we considered here.  The fractional change in the number of
particles produced due to finite volume effects is expected to be of
order $1/(K_{\rm heal}R)=\xi/(2\pi R)$=(cutoff
wavelength)/($2\pi\times$size of the condensate).  The ratio between
the healing length and the BEC Thomas--Fermi radius can be expressed
as a ratio between the harmonic trap length and the scattering length.
\begin{equation}
\frac{R}{\xi}=
\frac{2\sqrt{\pi}}{15^{1/10}}
\left(\frac{N_c a}{a_{ho}}\right)^{2/5}.
\end{equation}
For a harmonic oscillator length of about $10$ microns, $N_c \approx
10^{6}$ atoms and a scattering length of 1 nanometer one gets
$R\approx 17 \xi$.  For a scattering length ten times larger (easily
achievable with a Feshbach resonance and still compatible with $N_c
a^3\ll 1$) and $a_{ho}\approx 1\;\mu$m one would get $R\approx
100\xi$.  This implies that $\xi/(2\pi R)\leq 1 \%$ and hence finite
volume effects are negligible.
  
To conclude, our analysis suggest that it should be already possible
to observe the process of cosmological particle creation in BEC
analogue systems, by changing the scattering length from an initial
value of about 100~nm to a final value of about 1~nm on times scales
shorter than milliseconds but larger than tens of microseconds.
 
%----------------------------------------------------------------
\section*{Acknowledgements}
%----------------------------------------------------------------

Research by CB is supported by the EC under the Marie Curie contract
HPMF-CT-2001-01203. Research by SL is supported by the US NSF under
grant No. PHY98-00967 and by the University of Maryland. Research by
MV is supported by the Marsden Fund administered by the Royal Society
of New Zealand [RSNZ]. SL wishes to thank T.~Jacobson for his help and
support, B.-L. Hu, E. Calzetta, K.Burnett, Y. Castin, C. Clark and W.
Phillips for illuminating discussions, and E. Bolda, D. Mattingly, G.
Pupillo and A.~Roura for their interesting remarks.

%========================================================================
% When possible, the references have proper Spires citations attached.
% This is supposed to help the Spires staff in updating their database.
% Don't muck around with this unless you know what you are doing.
% Don't touch the commented ``citation = '' lines.
%========================================================================

%----------------------------------------------------------------
%----------------------------------------------------------------

\begin{thebibliography}{99}
%========================================================================
\bibitem{Birrell-Davies}
N.~D.~Birrell and P.~C.~Davies,
``Quantum Fields In Curved Space,''
Cambridge University Press, Cambridge 1982.

%------------------------------------------------------------------------
\bibitem{Vafa-Strominger}
A.~Strominger and C.~Vafa,
``Microscopic Origin of the Bekenstein-Hawking Entropy,''
Phys.\ Lett.\ B {\bf 379}, 99 (1996)
[arXiv:hep-th/9601029].
%%CITATION = HEP-TH 9601029;%%

%------------------------------------------------------------------------
\bibitem{Hu1}
B.~L.~Hu and E.~Verdaguer,
``Stochastic gravity: A primer with applications,''
Class.\ Quant.\ Grav.\  {\bf 20}, R1 (2003)
[arXiv:gr-qc/0211090].
%%CITATION = GR-QC 0211090;%%

%------------------------------------------------------------------------
\bibitem{Hu2}
B.~L.~Hu and E.~Verdaguer,
``Recent advances in stochastic gravity: Theory and issues,''
arXiv:gr-qc/0110092.
%%CITATION = GR-QC 0110092;%%

%------------------------------------------------------------------------
\bibitem{Parentani}
R.~Parentani,
``What did we learn from studying acoustic black holes?,''
Int.\ J.\ Mod.\ Phys.\ A {\bf 17}, 2721 (2002)
[arXiv:gr-qc/0204079].
%%CITATION = GR-QC 0204079;%%

%------------------------------------------------------------------------
\bibitem{Unruh}
W.~G.~Unruh,
``Notes On Black Hole Evaporation,''
Phys.\ Rev.\ D {\bf 14}, 870 (1976).
%%CITATION = PHRVA,D14,870;%%
\\
W. G. Unruh, 
``Experimental black hole evaporation?'', 
Phys. Rev. Lett, {\bf 46},  1351 (1981).
%%CITATION = NONE;%%

%------------------------------------------------------------------------
\bibitem{acoustic}
M.~Visser,
``Acoustic black holes: Horizons, ergospheres, and Hawking radiation,''
Class.\ Quant.\ Grav.\  {\bf 15} (1998) 1767
[arXiv:gr-qc/9712010].
%%CITATION = GR-QC 9712010;%%

%------------------------------------------------------------------------
\bibitem{Brandenberger}
J.~Martin and R.~H.~Brandenberger,
``The trans-Planckian problem of inflationary cosmology,''
Phys.\ Rev.\ D {\bf 63}, 123501 (2001)
[arXiv:hep-th/0005209].
%%CITATION = HEP-TH 0005209;%%

%------------------------------------------------------------------------
\bibitem{Garay1}
L.~J.~Garay, J.~R.~Anglin, J.~I.~Cirac and P.~Zoller,
``Black holes in Bose-Einstein condensates,''
Phys.\ Rev.\ Lett.\  {\bf 85}, 4643 (2000)
[arXiv:gr-qc/0002015].
%%CITATION = GR-QC 0002015;%%

%------------------------------------------------------------------
\bibitem{Garay2}
L.~J.~Garay, J.~R.~Anglin, J.~I.~Cirac and P.~Zoller,
``Sonic black holes in dilute Bose-Einstein condensates,''
Phys.\ Rev.\ A {\bf 63}, 023611 (2001)
[arXiv:gr-qc/0005131].
%%CITATION = GR-QC 0005131;%%

%------------------------------------------------------------------------
\bibitem{bec-cqg}
C.~Barcel\'o, S.~Liberati and M.~Visser,
``Analog gravity from Bose-Einstein condensates,''
Class.\ Quant.\ Grav.\  {\bf 18} (2001) 1137
[arXiv:gr-qc/0011026].
%%CITATION = GR-QC 0011026;%%

%------------------------------------------------------------------------
\bibitem{abh}
M.~Novello, M.~Visser and G.~Volovik,
``Artificial Black Holes'',
(World Scientific, Singapore, 2002).

%------------------------------------------------------------------------
\bibitem{laval}
C.~Barcel\'o, S.~Liberati and M.~Visser,
``Towards the observation of Hawking radiation in Bose-Einstein  condensates,''
Int.\ J.\ Mod.\ Phys.\ A, in press; %{\bf --}, ---- (2003)
arXiv:gr-qc/0110036.
%%CITATION = GR-QC 0110036;%%

%------------------------------------------------------------------------
\bibitem{grf}
C.~Barcel\'o, S.~Liberati and M.~Visser,
``Analogue models for FRW cosmologies'',
Int.\ J.\ Mod.\ Phys.\ D, in press; arXiv:gr-qc/0305061.
%%CITATION = GR-QC 0305061;%% 

%------------------------------------------------------------------------
\bibitem{Hu-Calzetta}
E.~A.~Calzetta and B.~L.~Hu,
``Bose-Novae as Squeezing of the Vacuum by Condensate Dynamics,''
arXiv:cond-mat/0208569.
%%CITATION = COND-MAT 0208569;%%
\\
E.~A.~Calzetta and B.~L.~Hu,
``Bose-Novae as Squeezing of Vacuum Fluctuations by Condensate Dynamics,''
arXiv:cond-mat/0207289.
%%CITATION = COND-MAT 0207289;%%

%------------------------------------------------------------------------
\bibitem{bosenova-exp} 
E. Donley \emph{et al.}, Nature {\bf 412} (2001) 295;
\\
N. R. Claussen, Ph.~D. Thesis, U.~of Colorado (2003).



%------------------------------------------------------------------------
\bibitem{Rb1}
S. L. Cornish, N. R. Claussen, J. L. Roberts, E. A. Cornell, C. E. Wieman,
``Stable 85Rb Bose-Einstein Condensates with Widely Tunable Interactions''
Phys. Rev. Lett {\bf 85}, 1795 (2000).
arXiv:cond-mat/0004290
\\
J. L. Roberts, N. R. Claussen, S. L. Cornish, E. A. Donley,
E. A. Cornell, C. E. Wieman, 
``Controlled Collapse of a Bose-Einstein Condensate''
arXiv:cond-mat/0102116
\\
Elizabeth A. Donley, Neil R. Claussen, Simon L. Cornish, Jacob L. Roberts, 
Eric A. Cornell, Carl E. Wieman,
``Dynamics of collapsing and exploding Bose-Einstein condensates''
arXiv:cond-mat/0105019

%------------------------------------------------------------------------
\bibitem{Griffin}
A. Griffin,
``Conserving and gapless approximations for an inhomogeneous Bose gas
at finite temperature''
Phys. Rev {\bf B53}, 9341 (1996).

%------------------------------------------------------------------------
\bibitem{Fischer}
P.~O.~Fedichev and U.~R.~Fischer,
``'Cosmological' particle production in oscillating ultracold Bose gases:  
The role of dimensionality,''
arXiv:cond-mat/0303063;
%%CITATION = COND-MAT 0303063;%%
\\
``Hawking radiation from sonic de Sitter horizons in expanding  
Bose-Einstein-condensed gases,''
arXiv:cond-mat/0304342;
%%CITATION = COND-MAT 0304342;%%
\\
``Observing quantum radiation from acoustic horizons in linearly expanding 
cigar-shaped Bose-Einstein condensates,''
arXiv:cond-mat/0307200
%%CITATION = COND-MAT 0307200;%%

%------------------------------------------------------------------------
\bibitem{Kagan-Castin}
Y.~Kagan, E.~L.~Surkov and G.~V.~Shlyapnikov,
``Evolution of a Bose-condensed gas under variations of the 
confining potential,'' 
Phys. Rev. {\bf A54}, R1753 (1996);
Y. Castin and R. Dum, 
``Bose-Einstein Condensates in Time Dependent Traps,''
Phys. Rev. Lett. {\bf 77}, 5315 (1996).

%------------------------------------------------------------------------
\bibitem{Kagan}
Y.~Kagan, E.~L.~Surkov and G.~V.~Shlyapnikov,
``Evolution and global collapse of trapped Bose condensates under 
variations of the scattering length,''
Phys.\ Rev.\ Lett.\  {\bf 79}, 2604 (1997)
[arXiv:physics/9705005].
%%CITATION = PHYS-ICS 9705005;%%

%------------------------------------------------------------------------
\bibitem{Volovik}
G.E. Volovik,
``Induced Gravity in Superfluid 3He,''
J.Low.Temp.Phys. {\bf 113}, 667 (1997)
[arXiv:cond-mat/9806010].

%------------------------------------------------------------------------
\bibitem{Castin}
Yvan Castin,
``Bose-Einstein condensates in atomic gases: simple theoretical results''
in 'Coherent atomic matter waves', Lecture Notes of Les Houches Summer
School, p.1-136, edited by R. Kaiser, C. Westbrook, and F. David, EDP
Sciences and Springer-Verlag (2001).

%------------------------------------------------------------------------
\bibitem{broken}
M.~Visser, C.~Barcelo and S.~Liberati,
``Acoustics in Bose-Einstein condensates as an example of broken 
Lorentz  symmetry,''
arXiv:hep-th/0109033.
%%CITATION = HEP-TH 0109033;%%

%------------------------------------------------------------------------
\bibitem{Kohler-Burnett} 
T. K\"ohler and K. Burnett, 
``Microscopic quantum dynamics approach to the dilute condensed Bose gas'',
Phys.\ Rev.\  {\bf A65}, 033601 (2002).

%------------------------------------------------------------------------
\bibitem{Gribakin93}
G.~F.~Gribakin and V.~V.~Flambaum,
``Calculation of the scattering length in atomic collisions using the semiclassical approximation'',
Phys.\ Rev.\  {\bf A48}, 546 (1993).

%------------------------------------------------------------------------
\bibitem{Weiner99}
J.~Weiner, V.~S.~Bagnato, S.~Zilio, and P.S.~Julienne,
``Experiments and theory in cold and ultracold collisions''
Rev.\ Mod.\ Phys.\ {\bf 71}, 1 (1999).

%------------------------------------------------------------------------
\bibitem{Williams99}
C.~J.~Williams {\etal}, 
``Determination of the scattering lengths of $^{39}K$ from $1_{u}$ photoassociation line shapes'',
Phys.\ Rev.\  {\bf A60}, 4427 (1999).
%------------------------------------------------------------------------
\bibitem{Bolda02}
E.~L.~Bolda, E.~Tiesinga, and P.~S.~Julienne,
``Effective-scattering-length model of ultracold atomic collisions and Feshbach resonances in tight harmonic traps''
Phys.\ Rev.\  {\bf A66}, 013403 (2002).

%------------------------------------------------------------------------
\bibitem{Parker}
L. Parker, 
``Thermal radiation produced by the expansion of the universe'', 
Nature {\bf 261}, 20 (1976).

%------------------------------------------------------------------------
\bibitem{Eckart}
C. Eckart,
``The penetration of a potential barrier by electrons''
Phys.\ Rev.\  {\bf 35}, 1303 (1930).

%------------------------------------------------------------------------
\bibitem{sono-I}
S.~Liberati, M.~Visser, F.~Belgiorno and D.~W.~Sciama,
``Sonoluminescence as a QED vacuum effect. I: The physical scenario,''
Phys.\ Rev.\ D {\bf 61}, 085023 (2000)
[arXiv:quant-ph/9904013].
%%CITATION = QUANT-PH 9904013;%% 

%------------------------------------------------------------------------
\bibitem{finite-volume}
S.~Liberati, M.~Visser, F.~Belgiorno and D.~W.~Sciama,
``Sonoluminescence as a QED vacuum effect. II: Finite volume effects,''
Phys.\ Rev.\ D {\bf 61} (2000) 085024
[arXiv:quant-ph/9905034].
%%CITATION = QUANT-PH 9905034;%%

%------------------------------------------------------------------------
\bibitem{bounds}
M.~Visser,
``Some general bounds for 1-D scattering,''
Phys.\ Rev.\ A {\bf 59} (1999) 427
[arXiv:quant-ph/9901030].
%%CITATION = QUANT-PH 9901030;%

%----------------------------------------------------------------
\end{thebibliography}
\end{document}